\documentclass[prb,amsmath,amssymb,twocolumn,showpacs,floatfix]{revtex4}
\usepackage{graphicx}
\usepackage{dcolumn}
\usepackage{bm}
\usepackage{subfigure}

\begin{document}

\title{Precision Search for Magnetic Order in the Pseudogap Regime of La$_{2-x}$Sr$_x$CuO$_4$ by Muon Spin Relaxation}

\author{W. Huang,$^1$}
\altaffiliation{Current address: Department of Physics and Astronomy, McMaster University, Hamilton, Ontario L8S 4M1, Canada}
\author{V. Pacradouni,$^1$}
\altaffiliation{Current address: Powertech Labs Inc., Surrey, British Columbia V3W 7R7, Canada} 
\author{M.P. Kennett,$^1$ S. Komiya,$^2$ and J.E. Sonier$^{1,3}$}
\affiliation{$^1$ Department of Physics, Simon Fraser University, Burnaby, British Columbia V5A 1S6, Canada\\
$^2$ Central Research Institute of Electric Power Industry, Yokosuka, Kanagawa 240-0196, Japan\\
$^3$ Canadian Institute for Advanced Research, Toronto, Ontario, M5G 1Z8, Canada}

\date{\today}
\begin{abstract}
We report a high precision search for orbital-like magnetic order in the pseudogap region of La$_{2-x}$Sr$_x$CuO$_4$ single crystals 
using zero-field muon spin relaxation (ZF-$\mu$SR). In contrast to previous studies of this kind, the effects of the 
dipolar and quadrupolar interactions of the muon with nearby nuclei are calculated. ZF-$\mu$SR spectra with a 
high number of counts were also recorded to determine whether a magnetically ordered phase exists in dilute 
regions of the sample. Despite these efforts, we find no evidence for static magnetic order of any kind in the 
pseudogap region above the hole-doping concentration $p \! = \! 0.13$.
\end{abstract}


\maketitle
A distinctive feature of high transition temperature ($T_c$) cuprate superconductors is the pseudogap region
that exists above $T_c$ and over a wide range of doping. For some time there has been much debate on whether
the pseudogap is a manifestation of a phase transition. Recently it has been demonstrated that the onset
of the pseudogap in optimally-doped Pb$_{0.55}$Bi$_{1.5}$Sr$_{1.6}$La$_{0.4}$CuO$_{6 + \delta}$
at a temperature $T^\ast \! \approx \! 3.5 T_c$ is likely a sign of a phase transition to a non-superconducting 
state with a broken symmetry.\cite{He:11} Although this is not necessarily a magnetically ordered state, 
proposed orders for the pseudogap state include time-reversal symmetry breaking phases that contain ordered 
circulating orbital currents, which either break~\cite{Chakravarty:01,Tewari:08} or preserve~\cite{Varma:06} translational symmetry. 
The strongest experimental evidence for an orbital-current phase are the 
the observations of an unusual translational-symmetry preserving magnetic order in YBa$_2$Cu$_3$O$_y$ 
and HgBa$_2$CuO$_{4+\delta}$ by spin-polarized neutron diffraction,\cite{Fauque:06,Mook:08,Li:08,Baledent:11,Li:11} 
which bear some resemblance to the ordered $\Theta_{II}$ circulating-current phase 
proposed in Ref.~\onlinecite{Varma:06}. It is worth pointing out, however, that by extrapolation the onset of this 
orbital-like magnetic order is expected to occur near $T_c$ at optimal doping, and hence its relationship to the phase transition reported 
in Ref.~\onlinecite{He:11} is unclear. Furthermore, to date such orbital-like magnetic order has not been
observed in any cuprate beyond a hole doping of $p \! = \! 0.135$, whereas the $\Theta_{II}$ phase is
predicted to persist up to $p \! \sim \! 0.19$. 

In contrast to the experimental techniques used in the above studies, local probes of magnetism, 
such as nuclear magnetic resonance (NMR) and zero-field muon spin relaxation (ZF-$\mu$SR) provide information on 
the magnetic volume fraction. Unfortunately, $^{89}$Y NMR experiments on Y$_2$Ba$_4$Cu$_7$O$_{15-\delta}$,
\cite{Strassle:08} Zeeman perturbed nuclear quadrupole resonance (NQR) measurements of 
YBa$_2$Cu$_4$O$_8$,\cite{Strassle:10} and ZF-$\mu$SR experiments on YBa$_2$Cu$_3$O$_y$ (Ref.~\onlinecite{Sonier:02}, \onlinecite{Sonier:09}) 
and La$_{2-x}$Sr$_x$CuO$_4$ (Ref.~\onlinecite{MacDougall:08}) have found no evidence for the onset of magnetic order at the 
pseudogap temperature $T^\ast$. One exception is the finding of anomalous magnetic order by ZF-$\mu$SR \cite{Sonier:09} 
in the same YBa$_2$Cu$_3$O$_{6.6}$ single crystal studied in Ref.~\onlinecite{Mook:08}. The magnetic order is 
characterized by an onset temperature and an average local dipolar magnetic field that are in quantitative agreement with 
the orbital-like magnetic order detected by polarized neutron diffraction. 
Yet the ZF-$\mu$SR measurements clearly show this form of magnetic order existing in only about $3\%$ of 
the sample, suggesting that it is associated with a minority phase in lower quality samples. 

Orbital-like magnetic order has also been observed in La$_{1.915}$Sr$_{0.085}$CuO$_4$ by polarized neutron diffraction,
\cite{Baledent:10} but is less pronounced than the long-range magnetic order that has been reported in the pseudogap regions of 
YBa$_2$Cu$_3$O$_y$ (Ref.~\onlinecite{Fauque:06}, \onlinecite{Mook:08}) and HgBa$_2$CuO$_{4+\delta}$ (Ref.~\onlinecite{Li:08}). In particular,
the magnetic order in La$_{1.915}$Sr$_{0.085}$CuO$_4$ is short range, two (rather than three) dimensional, and
occurs at a temperature far below $T^\ast$. Yet no such magnetic order was observed in a ZF-$\mu$SR study 
of $x \! \ge \! 0.13$ samples by MacDougall {\it et al.}\cite{MacDougall:08} 
Nevertheless, the subtle nature of the magnetic order observed in La$_{1.915}$Sr$_{0.085}$CuO$_4$ warrants a more
precise ZF-$\mu$SR search to verify its existence and determine
whether it is an intrinsic property of the pseudogap phase of La$_{2-x}$Sr$_x$CuO$_4$. 

Here we report two significant advances in the application of ZF-$\mu$SR to search for 
magnetic order in the pseudogap region of La$_{2-x}$Sr$_x$CuO$_4$. First, we accurately
determine the interactions of the positive muon ($\mu^+$) with the nuclear spin system, allowing us to identify any 
residual relaxation of the ZF-$\mu$SR spectrum that could be ascribed to static magnetic order.
In doing so we have accurately identified the muon stopping site. Second, we have acquired ZF-$\mu$SR spectra
of higher statistics than in previous works, enabling a search for dilute or short-range magnetic order.

\section{Experimental Details}
The experiments were performed on platelet-like single crystals of La$_{2-x}$Sr$_x$CuO$_4$ cut from a 
travelling-solvent floating zone (TSFZ) growth rod. The TSFZ growth procedure that was followed 
is decribed elsewhere.\cite{Komiya:02} The single crystals cut from the TSFZ rod were annealed 
at 800 to 900 $^\circ$C in an oxygen partial pressure to remove oxygen defects in accordance with 
the oxygen nonstoichiometry of La$_{2-x}$Sr$_x$CuO$_4$.\cite{Kanai:97}
Strontium concentrations greater than $x \! = \! 0.125$ were chosen to ensure the absence of 
static antiferromagnetism or the spin-glass-like magnetism previously observed by ZF-$\mu$SR in lower doped samples.\cite{Niedermayer:98,Panagopoulos:02}  
Magnetic susceptibility measurements of the bulk superconducting transition temperature by a superconducting 
quantum interference device (SQUID) yield $T_c \! = \! 37.6, 37.3, 28$ and 17 K for the $x \! = \! 0.15$, 0.166, 0.216, 
and 0.24 samples, respectively. 

The ZF-$\mu$SR measurements of the La$_{2-x}$Sr$_x$CuO$_4$ single crystals were performed on 
the M15 and M20B surface muon beam lines at TRIUMF. Positive muons implanted into the sample Larmor precess about the local field $B$ and decay according to $\mu^+ \rightarrow e^+ + \nu_e + \bar{\nu}_\mu$, with a mean life time $\tau \! \sim \! 2.2~\mu$s. The ZF-$\mu$SR signal is generated from detection of the decay positrons, which are preferentially emitted along the muon spin direction. 
The samples were mounted with the crystallographic $c$-axis of the La$_{2-x}$Sr$_x$CuO$_4$ single crystals parallel to the muon beam momentum. 
The initial muon spin polarization ${\bf P}(0)$ was oriented perpendicular to the $c$-axis using a Wien filter. This has the advantage that neither positron detector directly faces the incoming muon beam. In this geometry the ZF-$\mu$SR ``asymmetry'' spectrum is defined as the difference between the number of decay positrons sensed by scintillator detectors positioned above (A) and below (B) the sample, divided by the sum of the counts in these two detectors

\begin{equation}
\frac{N_A - N_B}{N_A + N_B} \equiv A(t) = aP(t)\, ,
\label{eq:asymmetry}
\end{equation}
where $a \! < \! 1/3$ is the initial asymmetry (dependent on the energy of the decay positrons and several experimental factors) and $P(t)$ is the time evolution of the muon spin polarization. The latter is modeled by an appropriate relaxation function $G(t)$
\begin{equation}
P(t) = G(t)\cos(\gamma_\mu Bt) \, ,
\label{eq:relaxation}
\end{equation}
where $\gamma_\mu$ is the muon gyromagnetic ratio and $B$ is the average internal magnetic field sensed by the muon.
Note that $B \! = \! 0$ in the absence of magnetic order. 

\section{Experimental Results}
In previous studies of La$_{2-x}$Sr$_x$CuO$_4$ the contribution of the nuclear moments to the ZF-$\mu$SR 
signal was assumed to be described by a static Gaussian Kubo-Toyabe (KT) relaxation function~\cite{Kubo:67,Kubo:81}

\begin{figure}
\centering
\includegraphics[width=9.0cm]{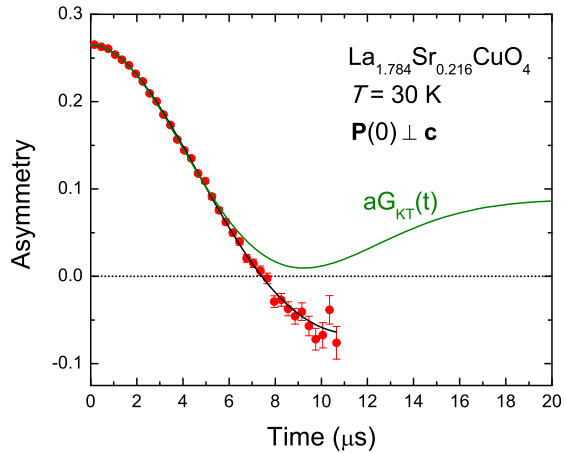}
\caption{(Color online) ZF-$\mu$SR signal of single crystal La$_{1.784}$Sr$_{0.216}$CuO$_4$ at $T \! = \! 30~$K recorded with the 
initial muon spin polarization {\bf P}(0) perpendicular to the $c$-axis. The solid green curve is a fit of the data 
below $t \! = \! 6~\mu$s to Eq.~(\ref{eq:KT}) multiplied by an asymmetry factor $a$. The solid black curve is a fit
to Eq.~(\ref{eq:fitfunction}).}
\label{fig:LSCO216}
\end{figure} 

\begin{equation}
G_{\rm KT}(t) = \frac{1}{3} + \frac{2}{3} (1- \Delta^2 t^2)\exp \left[ -\frac{1}{2} \Delta^2 t^2 \right] \, ,
\label{eq:KT}
\end{equation}

\noindent where $\Delta$ is the second moment of the local magnetic field distribution at the muon site. 
The relaxation function $G_{\rm KT}(t)$ assumes that the muon is immobile, the local magnetic fields acting on 
the muon spin are static, and that the local magnetic field distribution is isotropic and Gaussian. 
The nuclear dipolar fields sensed by the positive muon $\mu^+$ are usually static, because correlation times of the nuclear moments are generally much longer than the muon life time. Assuming that the nuclear dipolar fields acting on the muon spin are randomly oriented and that each muon sees a unique local field over the duration of its life time, 
1/3 of the muon spins will be parallel to the field and not evolve in time. Consequently, the recovery of the muon spin polarization to 1/3 of its initial value  at late times is characteristic of static fields. However, in a single crystal the contribution of the surrounding nuclear moments can substantially deviate from the random field approximation, with no recovery of $P(t)$ to 1/3. 
For example, Fig.~\ref{fig:LSCO216} shows that the static Gaussian Kubo-Toyabe function does not fully describe the ZF-$\mu$SR spectrum of single crystal 
La$_{1.784}$Sr$_{0.216}$CuO$_4$ for the case where {\bf P}(0) is perpendicular to the $c$-axis --- despite being in a region of the phase diagram where only the
nuclear dipole moments are expected to contribute to the ZF relaxation function. Of particular note, 
the muon spin polarization dips below zero, whereas $G_{\rm KT}(t)$ does not. To circumvent this problem the ZF-$\mu$SR signals for 
La$_{2-x}$Sr$_x$CuO$_4$ in Ref.~\onlinecite{MacDougall:08} were truncated at $t \! = \! 8~\mu$s to facilitate fits to Eq.~(\ref{eq:KT}).

\begin{figure}
\centering
\includegraphics[width=9.0cm]{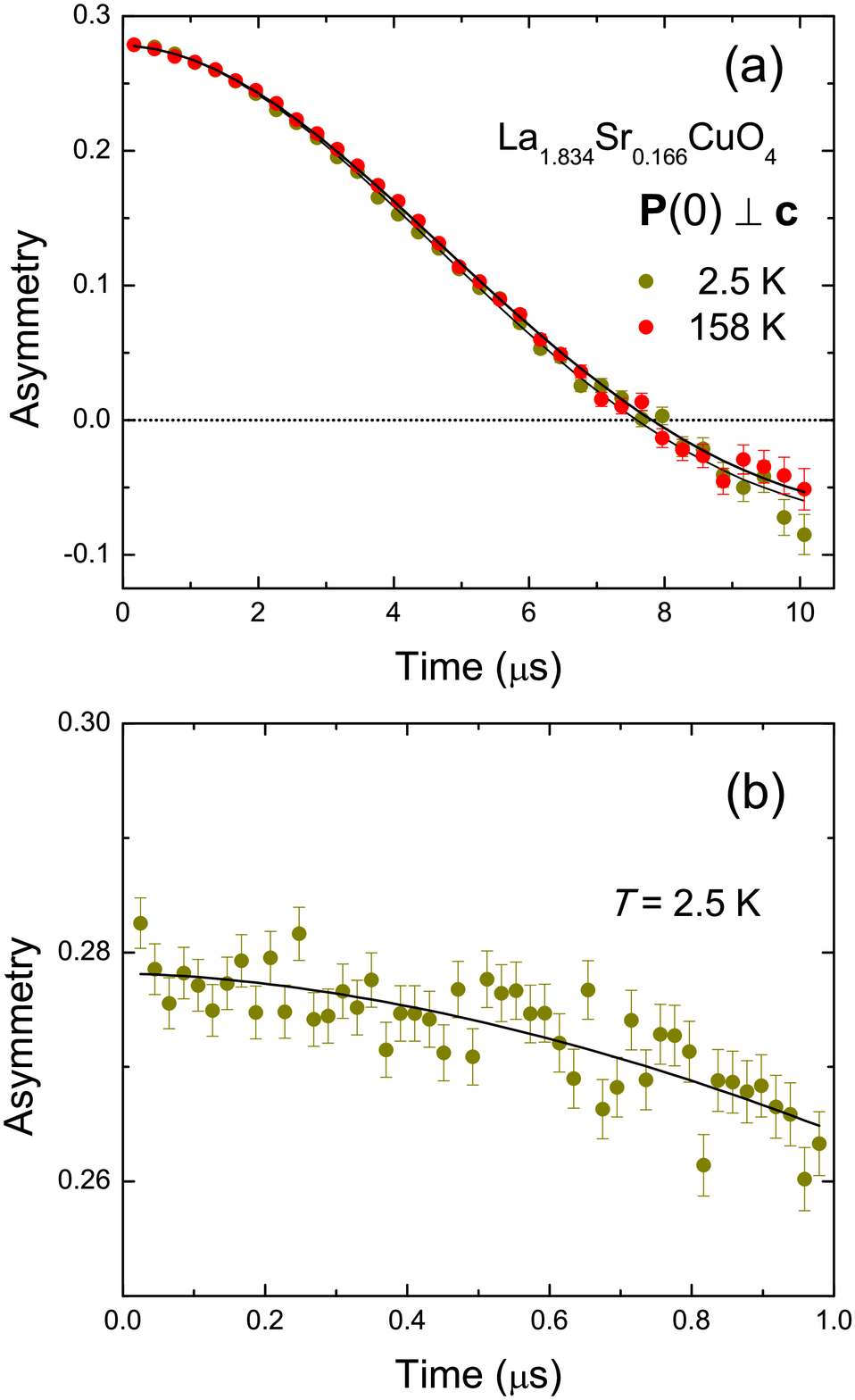}
\caption{(Color online) (a) Comparison of the ZF-$\mu$SR signals of La$_{1.834}$Sr$_{0.166}$CuO$_4$ at $T \! = \! 2.5$~K and $T \! = \! 158$~K measured over a 10 $\mu$s time range
with {\bf P}(0) perpendicular to the $c$-axis. 
(b) The early time ZF-$\mu$SR signal of La$_{1.834}$Sr$_{0.166}$CuO$_4$ at $T \! = \! 2.5$ K shown over the first 1 $\mu$s. In both figures the solid curves are 
fits to the relaxation function of Eq.~(\ref{eq:fitfunction}).}
\label{fig:FigAsy}
\end{figure} 

\begin{figure}
\centering
\includegraphics[width=9.0cm]{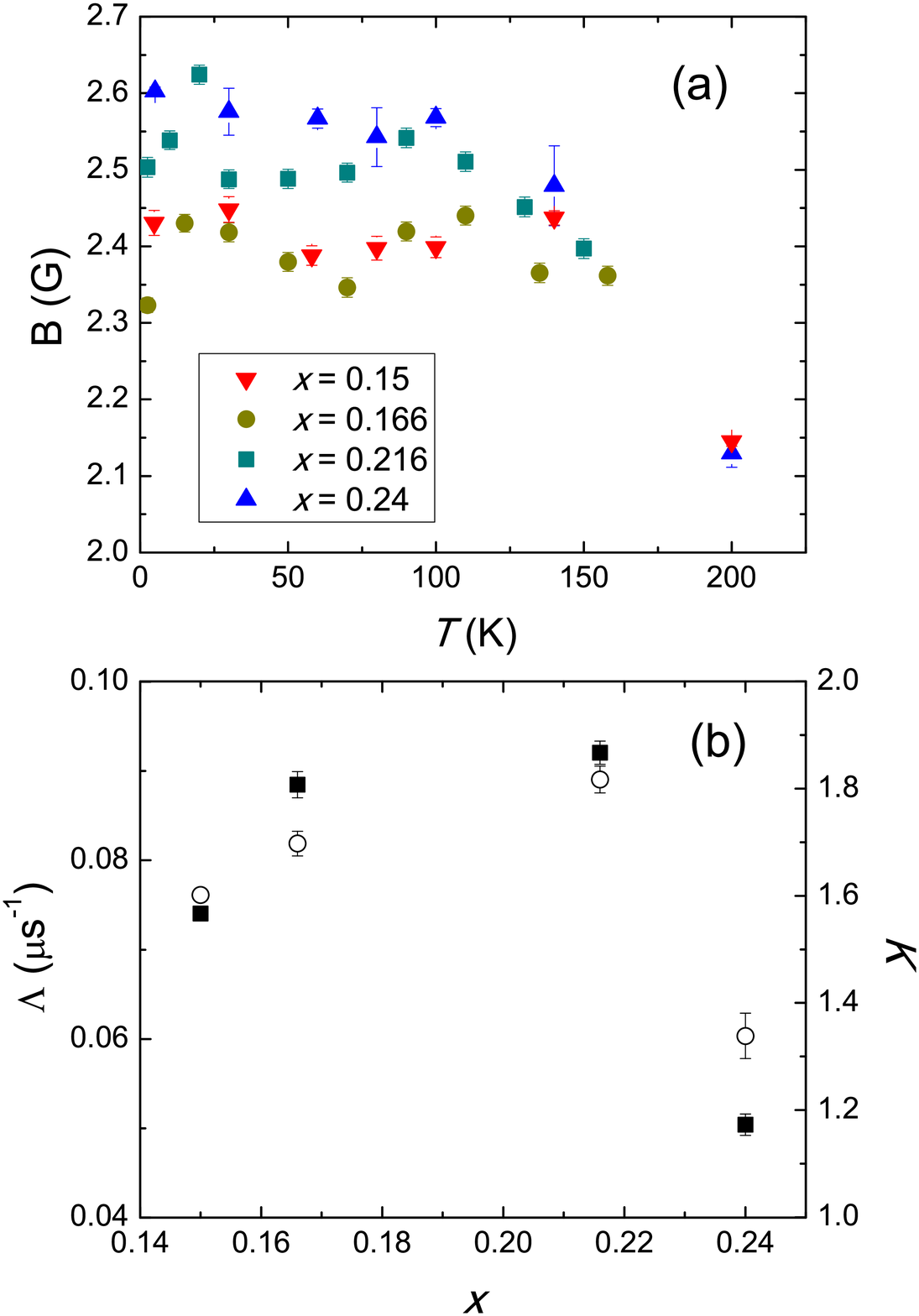}
\caption{(Color online) Results of fits to Eq.~(\ref{eq:fitfunction}). (a) Temperature dependence of $B$ for all four samples. (b) Dependence of the relaxation rate $\Lambda$ (solid squares) and the exponent $K$ (open circles) on Sr content $x$. Note, both $\Lambda$ and $K$ are temperature independent fit parameters.}
\label{fig:FigRlxPow}
\end{figure}

To more accurately portray changes in the functional form of the ZF-$\mu$SR signal as a function of temperature and Sr content $x$ ({\it i.e.} hole doping concentration), 
here we fit the ZF-$\mu$SR spectra to a simple phenomenological power-exponential relaxation function, such that
\begin{equation}
P(t) = \exp \left[-(\Lambda t)^K \right] \cos[\gamma_\mu B(T) t] \, ,
\label{eq:fitfunction}
\end{equation}
\noindent with the relaxation rate $\Lambda$ and the exponent $K$ treated as temperature independent quantities, 
and the average local magnetic field $B$ being the only fit parameter allowed to vary with temperature. 
Figure~\ref{fig:FigAsy}(a) shows representative time spectra for La$_{1.834}$Sr$_{0.166}$CuO$_4$ 
that are well described by Eq.~(\ref{eq:fitfunction}) over the entire 10 $\mu$s time range.
A fit to Eq.~(\ref{eq:fitfunction}) is also shown for La$_{1.784}$Sr$_{0.216}$CuO$_4$ in Fig.~\ref{fig:LSCO216}.  
The evolution of the ZF-$\mu$SR signal with temperature is reflected in the temperature dependence of $B$ 
shown in Fig.~\ref{fig:FigRlxPow}(a). Consistent with the findings of MacDougall {\it et al.},\cite{MacDougall:08} 
the ZF-$\mu$SR signal does not exhibit a temperature dependence characteristic of a magnetic phase transition in any 
of the samples. At $T \! = \! 200$ K, the value of $B$ is clearly reduced, but by the same amount at $x \! = \! 0.15$ 
and $x \! = \! 0.24$. Hence, the reduction of $B$ at high temperatures is likely caused by muon diffusion, whereby the mobile 
$\mu^+$ sees a time-averaged field over its life time. This same conclusion was reached in Ref.~\onlinecite{MacDougall:08}.

The dependence of $\Lambda$ and $K$ on the Sr content $x$ is shown in Fig.~\ref{fig:FigRlxPow}(b). There is some increase in the values of both parameters
with increasing $x$, but a clear reduction of $\Lambda$ and $K$ for the $x \! = \! 0.24$ sample. The former behavior may be the result of antiferromagnetic
fluctuations, dilute regions of static magnetism and/or the relaxation caused by the Sr nuclei. While the precise source is unclear, there is
no evidence for static magnetic order in any of these samples. The smaller values of $\Lambda$ and $K$ at $x \! = \! 0.24$ indicate another contribution
to the ZF-$\mu$SR signal. This is likely paramagnetic moments that are known to be present in heavily-overdoped 
La$_{2-x}$Sr$_x$CuO$_4$ above $x \! \sim \! 0.19$.\cite{Oda:91,Nakano:94,Wakimoto:05,Torrance:89} The onset of the 
Curie-like paramagnetism is the probable cause of the slightly larger values of $B$ at $x \! = \! 0.216$ and 
$x \! = \! 0.24$ in Fig.~\ref{fig:FigRlxPow}(a). 
        
In Ref.~\onlinecite{Sonier:09}, three distinct magnetic components were observed in the ZF-$\mu$SR signal of a large YBa$_2$Cu$_3$O$_{6.6}$ single crystal. Neutron scattering experiments on this same single cystal show that the sample contains magnetic order associated with the impurity ``green phase'' Y$_2$BaCuO$_5$,\cite{Mook:010204} 
and two additional kinds of unusual magnetic order.\cite{Mook:010204,Mook:08} While one of the latter two magnetic components is manifested as a slow 
relaxing component that is also observed in higher quality samples, the other two forms of magnetic order are discernible as small-amplitude, rapidly-damped 
oscillatory components in the early time range of the ZF-$\mu$SR signal. The small amplitudes indicate that the magnetic orders are confined to small volume fractions of 
the YBa$_2$Cu$_3$O$_{6.6}$ single crystal. However, Fig.~\ref{fig:FigAsy}(b) shows that there are no such oscillatory components in the early time 
ZF-$\mu$SR signal of La$_{1.834}$Sr$_{0.166}$CuO$_4$. We have also examined the early time spectra of the other samples at various temperatures above
$T \! = \! 2.3$~K, and likewise find no evidence for any kind of short-range or dilute magnetic order. 

Next we calculate the contribution of the nuclei to the polarization function $P(t)$, 
in an effort to fully account for the observed ZF-$\mu$SR spectrum. 

\section{Calculation of Nuclear-Induced Relaxation}
Here we describe a general numerical method for calculating the ZF relaxation function resulting from the dipolar magnetic and quadrupolar electrostatic 
interactions of the $\mu^+$ with an arbitrary number of neighboring nuclei. We consider only the interactions of the nuclei with the muon, and ignore
interactions amongst the nuclei themselves. This approximation is justified by the size of $\gamma_\mu$, which is about an order of 
magnitude larger than the gyromagnetic ratios of the nuclei in La$_{2-x}$Sr$_x$CuO$_4$. We also assume that the electric field gradient (EFG) 
at each nuclear site is due to the Coulomb field of the unscreened positively charged muon, and the non-symmetric charge 
distribution of the crystal itself ({\it i.e.} the crystal EFG). 

The interactions between the muon spin ($S \! = \! 1/2$) and $N$ surrounding nuclei of spin $I$, as well as the effect 
of the crystal EFG, is described by the following Hamiltonian \cite{Celio:86, Abragam:61}
\begin{equation}
H = \sum^{N}_{j = 1}(H^{D}_{j} + H^{Q}_{\mu j} + H^{Q}_{o j}) \, ,
\label{eq:Hamiltonian}
\end{equation}
where
\begin{equation}
H^{D}_{j} = \frac{ \hbar^2 \gamma_{\mu} \gamma_{j} }{ r^{3}_{j} } [ {\bf S} \cdot {\bf I}_j - 3( {\bf S} \cdot {\bf n}_j )( {\bf I}_j \cdot {\bf n}_j ) ] \, ,
\label{eq:dipolar}
\end{equation}
\begin{equation}
H^{Q}_{\mu j} = \hbar \omega^Q_{\mu j} [ ( {\bf I}_j \cdot {\bf n}_j ) \cdot ( {\bf I}_j \cdot {\bf n}_j ) - I(I+1)/3 ] \, ,
\label{eq:quadrupolar}
\end{equation}
and
\begin{equation}
H^{Q}_{o j} = \frac{\hbar \omega^Q_{o j}}{2} \left[ I^2_{jz} - \frac{1}{3} I_j (I_j +1) + \frac{1}{6} \eta (I^2_{j+} +I^2_{j-}) \right] \, .
\label{eq:crystalquadrupolar}
\end{equation}

In the above equations, the term $H^{D}_{j}$ is the dipole-dipole interaction between the positive muon and the $j^{\rm th}$
nucleus, $H^Q_{\mu j}$ is associated with the quadrupolar energy of the nuclear spin $I_j$ due to the EFG generated by 
the positive muon, ${\bf n}_j$ is the unit vector pointing in the direction along the straight line that connects the 
muon to the $j^{\rm th}$ nucleus located a distance $r_j$ away, and $\gamma_{\mu}$ and $\gamma_j$ are the gyromagnetic 
ratios of the muon and nuclei, respectively. The quadrupolar coupling constant $\omega^Q_{\mu j}$ is proportional 
to $1/r_j^3$.\cite{Schenck:85} The term $H^{Q}_{o j}$ represents the quadrupolar energy of the nuclear spin due to the 
crystal EFG, with a quadrupolar coupling constant $\omega^Q_{o j}$. The constant $\eta$ is an asymmetry parameter 
which specifies the symmetry of the crystal EFG around the nucleus.\cite{Abragam:61} In our calculation, the
values of the quadrupolar coupling $\omega^Q_{o j}$ and the asymmetry parameter $\eta$ for La and Cu are taken
from the literature.
In particular, $\omega^Q_{o j} \! = \! 2\pi \nu_Q$ with $\nu_Q \! = \! 34.0,~31.0$ and 6.40 Hz for $^{63}$Cu, $^{65}$Cu and $^{139}$La, 
respectively.\cite{Hunt:01,MacLaughlin:94} At Cu sites, $\eta \! = \! 0.03$,\cite{Tsuda:88} while at La sites, $\eta \! = \! 0.02$.\cite{MacLaughlin:94}
In addition, the weighted averages of the two isotopes
$^{63}$Cu and $^{65}$Cu are used for the gyromagnetic ratio $\gamma_{\rm Cu}$ and the nuclear quadrupole moment $Q_{\rm Cu}$ of the Cu nuclei.  

Since the only stable isotope of Sr with nonzero spin is $^{87}$Sr with a natural abundance of 7$\%$, 
the Sr nuclei can be neglected in the calculations. We note that this is consistent with the study by MacDougall {\it et al.},\cite{MacDougall:08} 
which showed that the relaxation rate of the ZF-$\mu$SR signal of La$_{2-x}$Sr$_x$CuO$_4$ varies little in the range $0.13 \! \leq \! x \! \leq \! 0.30$.
Consequently, it is sufficient to perform our calculations for a positive muon residing in a single crystal 
of the parent compound La$_2$CuO$_4$.
 
\begin{figure}
\centering
\includegraphics[width=8.0cm]{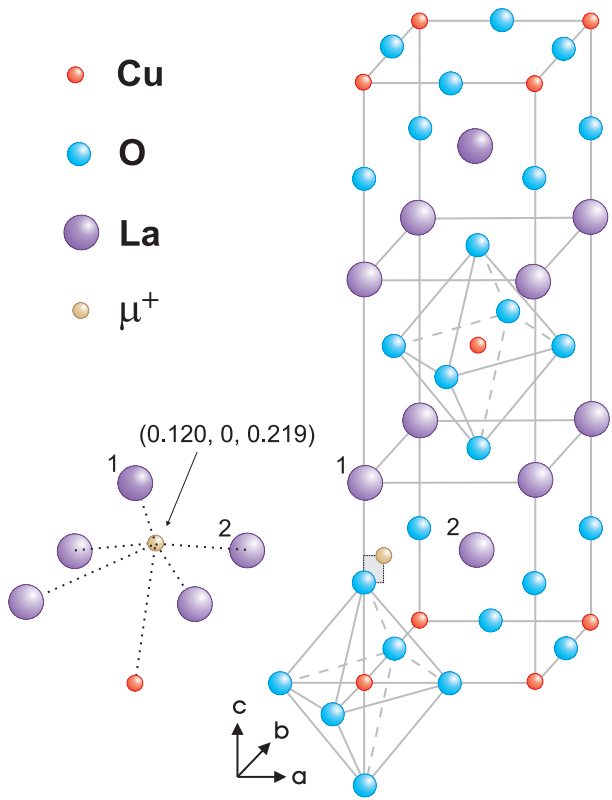}
\caption{(Color online) The muon D site (0.120, 0, 0.219) in the tetragonal unit cell of La$_2$CuO$_4$, which best describes the ZF-$\mu$SR signal of
La$_{2-x}$Sr$_x$CuO$_4$ at $x \! \ge \! 0.13$ (see Fig.~\ref{fig:x166x13}). The coordinate values are in multiples of their 
respective lattice constants $a \! = \! 3.80$~\AA, $b \! = \! 3.80$~\AA~ and $c \! = \! 13.12$~\AA.\cite{Radovic:08} 
The small shaded rectangle visually emphasizes that the muon at the D site near the apical O resides on the $a$-$c$ face of the unit cell.
The lower left of the picture shows the 5 La nuclei and the 1 Cu nucleus included in the calculation of $P(t)$ 
for this muon site (Two of the La atoms are labeled 1 and 2 to identify them in the full crystal structure).}
\label{fig:crystal}
\end{figure}

With knowledge of the Hamiltonian [Eq.~(\ref{eq:Hamiltonian})], the time dependence of the muon spin polarization may be calculated
from the density matrix of the spin system as
\begin{equation}
P(t) = {\rm Tr} \left\{ \rho(0) e^{ \frac{iHt}{\hbar} } \left[ \sigma_{\mu} \otimes \left( \otimes^{N}_{j=1} {\bf 1}_{D_j} \right) \right] e^{ \frac{-iHt}{\hbar} } \right\} \, .
\label{eq:timedepend}
\end{equation}
Here $\sigma_{\mu} = {\bf \sigma}_{\mu} \cdot {\bf P}(0) = \sigma_x\sin\theta \cos\beta + \sigma_y\sin\theta \sin\beta + \sigma_z\cos\theta$
is the projection of the muon spin along the direction of the initial polarization {\bf P}(0), with $\theta$ and 
$\beta$ the polar and azimuthal angles, respectively, between ${\bf P}(0)$ and the crystallographic axes (see Fig.~\ref{fig:crystal}). 
Note that from Eq.~(\ref{eq:crystalquadrupolar}), the crystal EFG sets the quantization axis to be along the $c$-axis of La$_{2-x}$Sr$_x$CuO$_4$.
The initial density matrix is $\rho (0)$, $D_j \! = \! 2I_j + 1$ is the spin degeneracy for the $j^{th}$ nuclear spin, and
${\bf 1}_n$ is the $n \! \times \! n$ identity matrix. 

For a system in which the nuclear spins are randomly oriented, the initial density matrix is
\begin{equation}
\rho (0)= \frac{1}{D} ({\bf 1}_2 + \sigma_{\mu}) \otimes \left( \otimes^{N}_{j=1} {\bf 1}_{D_j} \right) \, ,
\label{eq:densitymatrix}
\end{equation}
where
\begin{equation}
D = 2 \prod^{N}_{j = 1} D_j \, ,
\end{equation} 
is the dimensionality of the Hamiltonian matrix. The time evolution of the muon spin polarization
$P(t)$ may be determined exactly by diagonalizing the $D \! \times \! D$ Hamiltonian. The main limitation of this approach is that it becomes
very computationally expensive as the number of nuclei (and hence $D$) increases. In this work the largest systems studied include 10 nuclei, corresponding
to $D \! = \! 2^{26} \! \approx \! 6.71 \! \times \! 10^7$. We found that for these large values of $D$, it was necessary to use an approximate method
to determine $P(t)$, and hence we used a method developed by Celio to study $P(t)$ in copper.\cite{Celio:86} This approximation method is based
on the Trotter formula and utilizes the random phase approximation. We verified good agreement with the exact calculation for some of
our calculations with less nuclei. 

\begin{figure}
\centering
\includegraphics[width=8.0cm]{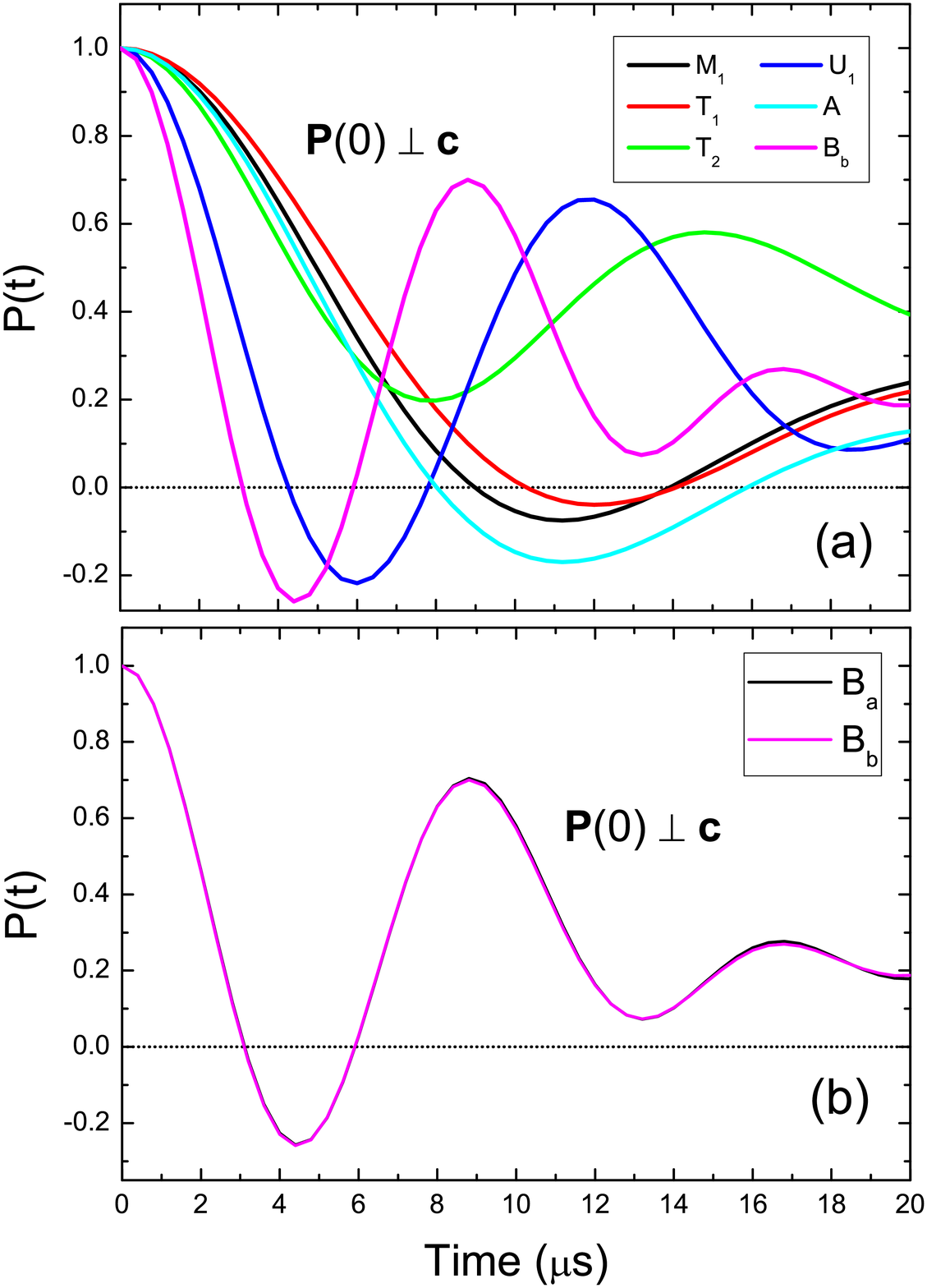}
\caption{(Color online) (a) Approximation method calculations of the muon spin polarization function $P(t)$ for different muon sites
and with the initial muon spin polarization {\bf P}(0) perpendicular to the $c$-axis and oriented at 45$^\circ$ with 
respect to the $a$-axis. The spatial coordinates of the
muon sites and the nuclei included in each calculation are shown in Table~\ref{tab:muonsites}. The sites M$_1$, T$_1$, T$_2$
and U$_1$ were previously considered in Ref.~\onlinecite{Sulaiman:94}. (b) $P(t)$ for the same muon site,
but with different nuclei used in the calculation (see Table~\ref{tab:muonsites}).} 
\label{fig:sitedependence}
\end{figure}
 
\begin{figure}
\centering
\includegraphics[width=9.0cm]{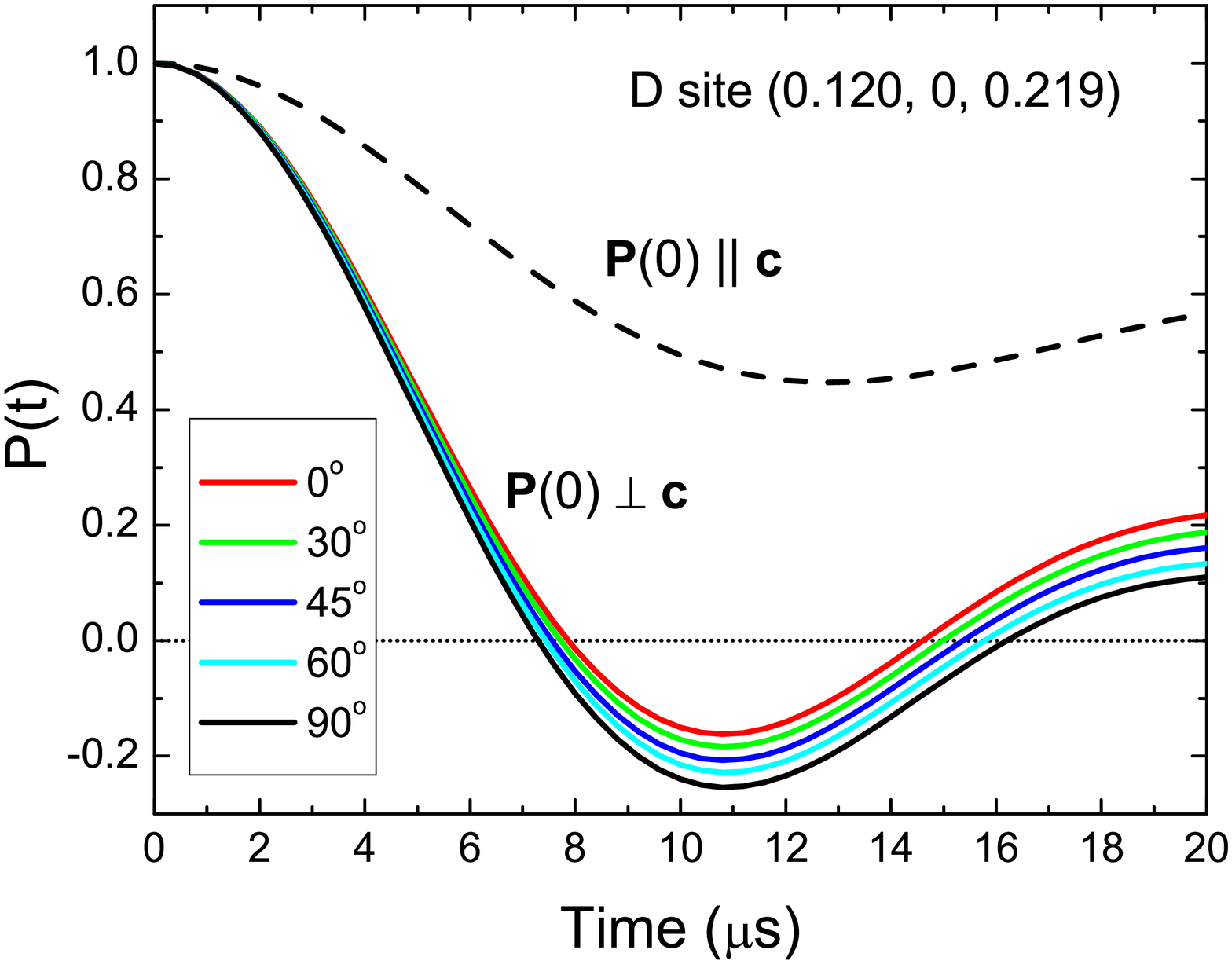}
\caption{(Color online) Time evolution of the  muon spin polarization $P(t)$ calculated for the muon D site of 
Fig.~\ref{fig:crystal} and different orientations of the initial muon spin polarization {\bf P}(0). 
The upper dashed curve shows $P(t)$ for {\bf P}(0) parallel to the $c$-axis.
The lower solid curves show $P(t)$ for different orientations of {\bf P}(0) in the $a$-$b$ plane, labeled
by the angle between {\bf P}(0) and the $a$-axis.}
\label{fig:angledependence}
\end{figure} 

The calculation of $P(t)$ is strongly dependent on the orientation of {\bf P}(0) with respect to the $c$-axis, the number and type of nuclei,
and the precise muon site. With the orientation of {\bf P}(0) fixed in the experiment, we used the approximation method to perform an exhaustive search 
for the function $P(t)$ that best describes the ZF-$\mu$SR signal of La$_{2-x}$Sr$_x$CuO$_4$ using numerous combinations 
of the muon site and the surrounding nuclei. Figure~\ref{fig:sitedependence}(a) shows $P(t)$ calculated by the approximation method for a handful 
of the numerous potential muon sites considered in our study. Some of these sites were considered in earlier
$\mu$SR studies.\cite{Hitti:91,Sulaiman:94} As demonstrated in Fig.~\ref{fig:sitedependence}(b),
only nearest-neighbor nuclei need to be included in the calculation of $P(t)$. In particular, note that both $B_{\rm a}$
and $B_{\rm b}$ correspond to the same muon site, but the calculation of $P(t)$ for $B_{\rm b}$ 
includes 3 additional Cu nuclei located further away from the muon (see Table~\ref{TableI}). The strong dependence of $P(t)$ 
on the muon site means that ZF-$\mu$SR can be used to accurately determine the location of the 
implanted $\mu^+$ in situations where there is a single muon site, the muon does not diffuse, and 
there are no additional sources of relaxation ({\it e.g.} electronic moments and/or loop-current order).

\begin{table}
\caption{\label{tab:muonsites} Muon sites of the polarization functions plotted 
in Fig.~\ref{fig:sitedependence}.
The spatial coordinates of the muon site are denoted by multiples of the lattice constants
$a$, $b$, and $c$ of La$_2$CuO$_4$. The nearest-neighbor nuclei used in the calculation
of $P(t)$ for each muon site are listed in the third column. Two Cu nuclei are denoted
as 2Cu, a La nucleus at site 1 (see Fig.~\ref{fig:crystal}) is denoted by La(1), etc.}
\begin{ruledtabular}
\begin{tabular}{lcl}
Label & Muon Site & Nuclei \\
\hline
M           &  (0.5, 0.0, 0.096)  & 2Cu, 2La(2) \\
T1          &  (0.2, 0.0, 0.15)   & 2Cu, La(1), 2La(2) \\
T2          &  (0.225, 0.0, 0.225)& 2Cu, La(1), 2La(2) \\
U1          &  (0.12, 0.0, 0.11)  & 2Cu, La(1), 2La(2) \\
A           &  (0.0, 0.0, 0.212)  & 5Cu, La(1), 4La(2) \\
B$_{\rm a}$ &  (0.1, 0.0, 0.1)    & Cu, La(1), 2La(2) \\
B$_{\rm b}$ &  (0.1, 0.0, 0.1)    & 4Cu, La(1), 2La(2) \\
D           &  (0.12, 0.0, 0.219) & Cu, La(1), 4La(2) \\
\end{tabular}
\end{ruledtabular}
\label{TableI}
\end{table}

\begin{figure}
\centering
\includegraphics[width=10.0cm]{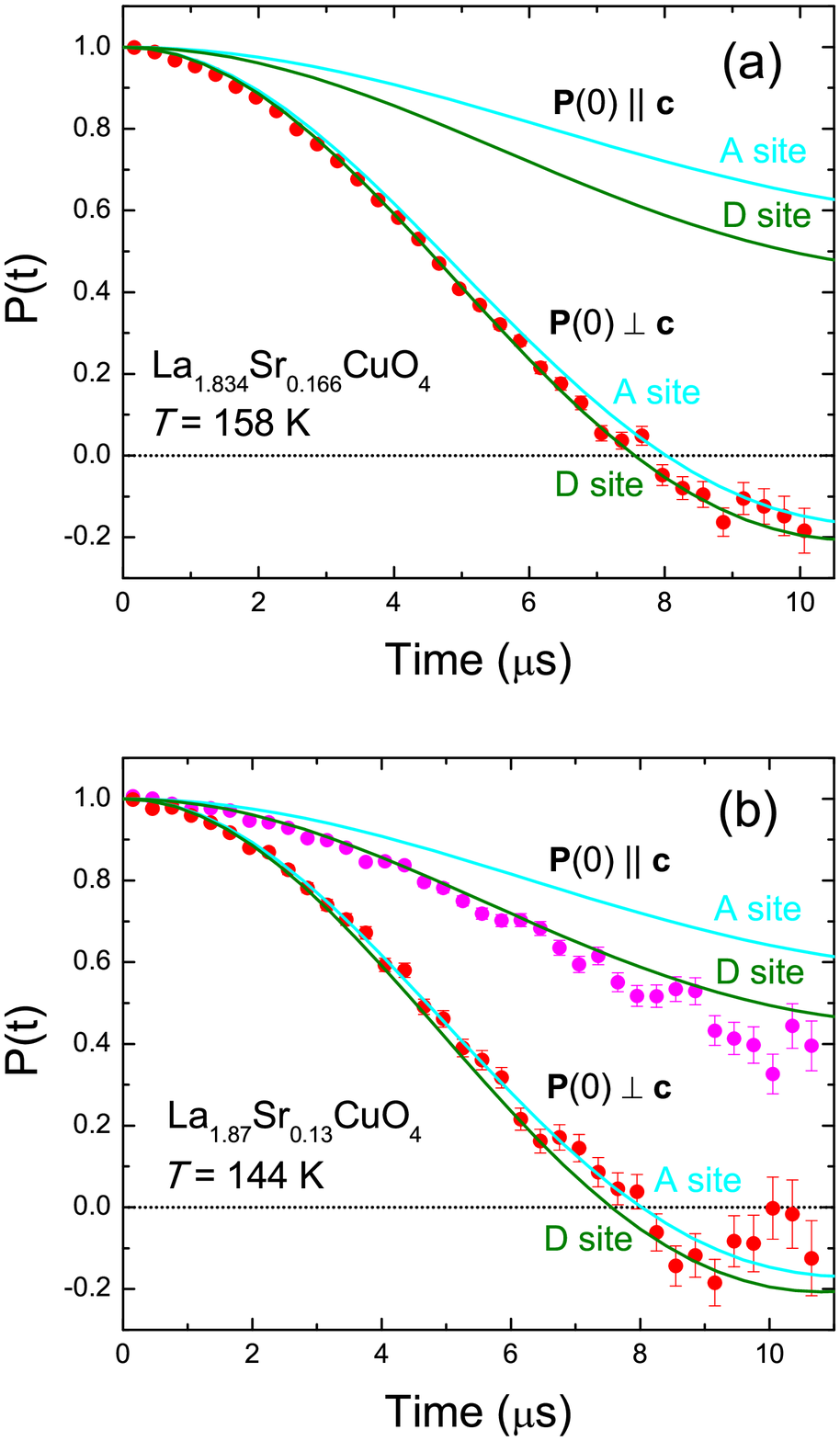}
\caption{(Color online) Comparisons between $P(t)$ calculated for the muon A and D sites and (a) the ZF-$\mu$SR signal
of La$_{1.834}$Sr$_{0.166}$CuO$_4$ at $T \! = \! 158$~K measured with {\bf P}(0) perpendicular to the $c$-axis, and (b) the
ZF-$\mu$SR signals of La$_{1.87}$Sr$_{0.13}$CuO$_4$ at $T \! = \! 144$~K from Ref.~\onlinecite{MacDougall:08}, for
{\bf P}(0) perpendicular to the $c$-axis, and {\bf P}(0) parallel to the $c$-axis. Note that the calculation for
${\bf P}(0) \! \perp \! {\bf c}$ assumes that the direction of {\bf P}(0) makes an angle of 45$^\circ$ with the $a$-axis.
Also, the ZF-$\mu$SR signals shown here have been divided by the initial asymmetry $a$.}
\label{fig:x166x13}
\end{figure}

Of all the muon sites we considered, only the polarization function 
$P(t)$ calculated for the site (0.120, 0, 0.219) shown in Fig.~\ref{fig:crystal} 
(which we refer to here as the D site) accounts for the observed ZF-$\mu$SR spectra (see Fig.~\ref{fig:x166x13}). This site is located approximately 
0.7~\AA~ from the apical oxygen, which agrees with the widely held view that the $\mu^+$ bonds to an oxygen atom in cuprates. While it is not exactly
one of the muon sites suggested in earlier works,\cite{Sulaiman:94,Hitti:91} it is consistent 
with ZF-$\mu$SR measurements of the antiferromagnetic phase of 
La$_2$CuO$_4$ by Hitti {\it et al.}\cite{Hitti:91} that restrict the muon site to the $a$-$c$ plane.

As shown in Fig.~\ref{fig:crystal} and indicated in Table~\ref{TableI}, five La nuclei of spin
$I \! = \! 7/2$ and one Cu nucleus of spin $I \! = \! 3/2$ are included in the calculation of $P(t)$ for the muon D site.
Figure~\ref{fig:angledependence} shows how $P(t)$ is dependent on the orientation of {\bf P}(0) with respect to
the crystal lattice. Although the orientation of {\bf P}(0) in the $a$-$b$ plane was random in our experiments, the 
ZF-$\mu$SR spectra were reproducible when the sample was rotated about the $c$-axis to a different position.
We can understand this as being a consequence of the equivalency of the sites (0.120, 0, 0.219) and (0, 0.120, 219) in
the tetragonal unit cell. Since these are occupied by the muon with equal probability, the ZF-$\mu$SR spectrum is an average 
of $P(t)$ for these two sites. For the case ${\bf P}(0) \! \perp \! {\bf c}$, at any orientation this superposition
is equivalent to $P(t)$ calculated with 
{\bf P}(0) lying in the $a$-$b$ plane and making an angle of $45^\circ$ with the $a$-axis.      
As shown in Fig.~\ref{fig:x166x13}(a), there is fairly good agreement between the calculation assuming the muon is located 
at the D site and the ZF-$\mu$SR spectrum of La$_{1.834}$Sr$_{0.166}$CuO$_4$ measured with ${\bf P}(0) \! \perp \! {\bf c}$.
This is also shown to be the case for the A site, where the muon is positioned directly above the apical O.
To distinguish between these two sites we consider the dependence on the angle between {\bf P}(0) and the $c$-axis.
In Fig.~\ref{fig:x166x13}(b) the calculations of $P(t)$ for both ${\bf P}(0) \! \perp \! {\bf c}$ and ${\bf P}(0) \! \parallel \! {\bf c}$ 
are compared to ZF-$\mu$SR spectra of La$_{1.87}$Sr$_{0.13}$CuO$_4$ from Ref.~\onlinecite{MacDougall:08}.
While this data is less accurate at late times due to lower muon counts, it is clear that the calculation for
the A site fails to describe the ZF-$\mu$SR signal with ${\bf P}(0) \! \parallel \! {\bf c}$.   

\section{Discussion and Conclusions}
\label{sec:Conclusion}
The ZF-$\mu$SR measurements presented here and in Ref.~\onlinecite{MacDougall:08} show no evidence for any form of 
static magnetism that can be directly linked to the pseudogap phase of La$_{2-x}$Sr$_x$CuO$_4$. Our determinaton
of the muon stopping site enables a good estimate of the magnitude of the average local magnetic field that should 
have been detected if static orbital-like magnetic order of the kind observed by spin-polarized 
neutron diffraction\cite{Fauque:06,Mook:08,Li:08,Baledent:11,Li:11,Baledent:10} were present. The neutron
results are in general agreement with the loop-current pattern of the $\Theta_{II}$ state proposed in 
Refs.~\onlinecite{Varma:06} and \onlinecite{Simon:02}, which consists of two oppositely circulating current 
loops per unit cell. If the orbital currents flow in the CuO$_2$ layers along the Cu-O and O-O bonds as originally
proposed, a $\mu^+$ residing at the D site (0.120, 0, 0.219) will experience a local field of 308~G/$\mu_B$.
For the maximum possible ordered magnetic moment of 0.02~$\mu_B$ deduced from the spin-polarized neutron measurements
of La$_{1.915}$Sr$_{0.085}$CuO$_4$,\cite{Baledent:10} the corresponding dipolar magnetic field sensed by the muon is 
6.2~G. In this case the damped ZF-$\mu$SR signal should oscillate with a period of $\sim \! 12$~$\mu$s,
such that one nearly complete oscillation is observed over the time range of our measurements.

The moments of the orbital-like magnetic order observed in hole-doped cuprates by spin-polarized
neutron diffraction are actually pointing at an angle of roughly $45^\circ$ with respect to the $c$-axis.
This is compatible with the orbital currents of the $\Theta_{II}$ phase flowing out and back into the CuO$_2$ 
layers through the apical oxygen atoms.\cite{Weber:09} In this scenario, the corresponding moments that are
perpendicular to the faces of the CuO$_6$ octahedra in La$_{2-x}$Sr$_x$CuO$_4$ point along directions making
an angle of $61^\circ$ away from the $c$-axis. For this arrangement with an ordered moment of 0.02~$\mu_B$, 
the average local field that a muon at the D site in La$_{2-x}$Sr$_x$CuO$_4$ would detect is about 90~G. 
Since 0.02~$\mu_B$ is the maximum possible value of the ordered moment observed in
La$_{1.915}$Sr$_{0.085}$CuO$_4$, and the orbital-like ordered moment in other cuprates\cite{Fauque:06,Mook:08,Li:08,Baledent:11,Li:11}
generally decreases with increased hole-doping, one might think of 90~G as an upper limit for the average local
field in the $x \! \ge \! 0.13$ samples considered here. On the other hand, a larger value is possible
if the ordered moment in La$_{1.915}$Sr$_{0.085}$CuO$_4$ is reduced due to a higher degree of disorder
and/or competition of the orbital-like magnetic order with the spin-density wave (SDW) order present in 
lower doped samples. Competition with SDW order has been suggested as a potential explanation for the 
smaller ordered moment observed in YBa$_2$Cu$_3$O$_{6.45}$ compared to that of YBa$_2$Cu$_3$O$_{6.5}$.\cite{Baledent:11}    
Regardless, the contribution of orbital-like magnetic order to the ZF-$\mu$SR signal will be strongly 
damped if it is short range as in the case of La$_{1.915}$Sr$_{0.085}$CuO$_4$. Yet no such component is
observed in any of the La$_{2-x}$Sr$_x$CuO$_4$ samples we measured.

The failure here and in Ref.~\onlinecite{MacDougall:08} to detect orbital-like magnetic order in
La$_{2-x}$Sr$_x$CuO$_4$ of the kind observed by spin-polarized neutron diffraction may indicate that the local fields 
are rapidly fluctuating oustide the $\mu$SR time window. It has also been suggested that the $\mu^+$ may
destroy loop-current order in cuprates,\cite{Shekhter:08} although thus far it has also eluded detection by NMR/NQR.
On the other hand, since the polarized neutron diffraction experiments cannot deduce magnetic volume fractions,
the orbital-like magnetic order could still be associated with a small minority phase that evolves
with hole doping.\cite{Sonier:09} Having said all of this, there is currently no disagreement between
the polarized neutron diffraction and ZF-$\mu$SR experiments on La$_{2-x}$Sr$_x$CuO$_4$, since there is
no overlap in the doping range of the samples studied by these two techniques. The ZF-$\mu$SR experiments have focussed 
on $x \! \geq \! 0.13$ samples to avoid major contributions from static SDW order or spin-glass-like magnetism.
Given that the orbital-like magnetic order observed by spin-polarized neutron diffraction in
La$_{1.915}$Sr$_{0.085}$CuO$_4$ is quite weak and has not been observed in other cuprates beyond $p \! = \! 0.135$,
it is conceivable that it is not present in $x \! \geq \! 0.13$ samples.
What can be said is that the ZF-$\mu$SR measurements of La$_{2-x}$Sr$_x$CuO$_4$ in the strontium concentration
(hole-doping) range $0.13 \! \leq x \! < \! 0.19$ do not support theoretically predicted loop-current phases, 
and hence favour an alternative explanation for the unusual magnetic order detected by spin-polarized 
neutron diffraction at lower hole doping. 

\section{Acknowledgements}
\label{Acknowledgements}
We thank the staff of TRIUMF's Centre for Molecular and Materials Science for technical assistance with our experiments, 
and G.M. Luke and G.J. MacDougall for sharing their data from Ref.~\onlinecite{MacDougall:08}. We also gratefully 
acknowledge the computing power provided by WestGrid, and thank Xiao-Li Huang for providing computing code used in the 
approximate calculations. This work was supported in part by the Natural Sciences and Engineering Research Council of Canada,
and the Canadian Institute for Advanced Research.

\end{document}